\documentstyle[12pt]{article}

\newcommand{\beq}{\begin{equation}}
\newcommand{\eeq}{\end{equation}}
\newcommand{\beqa}{\begin{eqnarray}}
\newcommand{\eeqa}{\end{eqnarray}}
\newcommand{\beqar}{\begin{eqnarray*}}
\newcommand{\eeqar}{\end{eqnarray*}}

\newcommand{\ka}{\kappa}

\renewcommand{\l}{\lambda}

\newcommand{\z}{\zeta}

\newcommand{\eg}{{\it e.g.,}\ }
\newcommand{\ie}{{\it i.e.,}\ }
\newcommand{\labell}[1]{\label{#1}} 
\newcommand{\reef}[1]{(\ref{#1})}
\newcommand\prt{\partial}

\newcommand\ls{\ell_s}
\newcommand\cF{{\cal F}}
\newcommand\cL{{\cal L}}

\newcommand\cO{{\cal O}}
\newcommand\tcO{{\tilde {{\cal O}}}}

\newcommand\hF{\hat{F}}
\newcommand\hA{\hat{A}}

\newcommand\hQ{\hat{Q}}
\newcommand\hP{\hat{\Phi}}
\newcommand\hS{\hat{S}}

\newcommand\hL{\hat{\cal L}}

\newcommand\tf{{\tilde f}}

\newcommand\tO{{\tilde O}}

\newcommand\Tr{{\rm Tr}}

\begin{document}

\thispagestyle{empty}
\rightline{\small hep-th/0105139 \hfill IPM/P-2001/009}
\vspace*{1cm}

\begin{center}
{\bf \Large 
 Dirac-Born-Infeld action, Seiberg-Witten map, \\[.25em]
and Wilson Lines }
\vspace*{1cm}

{Mohammad R. Garousi\footnote{E-mail:
garousi@theory.ipm.ac.ir}}\\
\vspace*{0.2cm}
{\it Department of Physics, University of Birjand, Birjand, Iran}\\
\vspace*{0.1cm}
and\\
{\it Institute for Studies in Theoretical Physics and Mathematics IPM} \\
{P.O. Box 19395-5531, Tehran, Iran}\\
\vspace*{0.4cm}

\vspace{2cm}
ABSTRACT
\end{center}
We write the recently conjectured action for transformation of 
the ordinary
Born-Infeld 
action 
 under the Seiberg-Witten
map  with one open Wilson contour in  a manifestly 
non-commutative gauge invariant form. This action contains the non-constant 
closed string fields, higher order derivatives  of the non-commutative 
gauge fields through the $*_N$-product, and  a  Wilson  operator. 
We  extend this 
non-commutative $D_9$-brane action to the action for $D_p$-brane  by 
transforming it  under T-duality. \\ 
 Using  this non-commutative $D_p$-brane  action we then evaluate  the 
linear couplings of the graviton and dilaton to the brane   for  arbitrary 
non-commutative parameters. By taking the Seiberg-Witten  limit we show 
that they reduce   exactly to the known results of the energy-momentum 
tensor of the  non-commutative super Yang-Mills theory. 
We take  this as 
an evidence that the non-commutative action in the Seiberg-Witten limit   
includes properly all  derivative correction terms.

\vfill
\setcounter{page}{0}
\setcounter{footnote}{0}
\newpage

\section{Introduction} \label{intro}

Recent years have seem dramatic progress in the understanding
of non-perturbative aspects of string theory -- see, \eg  \cite{excite}.
With these studies has come the realization that solitonic extended
objects, other than just strings, play an essential role.
An important object in these investigations has been Dirichlet
branes \cite{joep}. D-branes are non-perturbative states whose 
perturbative excitations are described by the fundamental open 
string states on their world-volumes. They are also source of  
 various perturbative closed string states  including the 
Ramond-Ramond states.

The massless excitations  of a single  D-brane in type II theories 
are a
U(1) vector, $A_a$,   a set of  scalars, $\Phi^i$, describing the transverse
motion  of the brane, and their superpartners\cite{leigh}.
The leading order low-energy effective action for the massless
fields corresponds to a dimensional reduction of a ten dimensional
U(1) Yang Mills theory. As usual in string theory, there are
higher order $\alpha'=\ls^2$ corrections,
where $\ls$ is the string length scale. As long as derivatives
of the field strength (and second derivatives of the scalars)
are small compared to $\ls$, then the action takes a Dirac-Born-Infeld
form \cite{bin}, in which the spacetime metric is the trivial 
Minkowskian metric and all other closed string fields are zero. 
One may naturally extend this flat space DBI action to the appropriate 
action in the curved space by adding  the non-constant closed string
fields, \ie
 the metric, dilaton and
Kalb-Ramond fields,  to the D-brane action.
The resulting 
 world-volume
action is\footnote{Our index conventions are that Greek indices take 
values in the entire ten-dimensional space-time, \eg $\mu,\nu=0,1,...,9$; 
early Latin indices take values in the world-volume , \eg $a,b=0,1,...,p$; 
and middle Latin indices take values in the transverse 
space, \eg $i,j=p+1,...,9$.},
\beqa
S&=&-T_p \int d^{p+1}x\, e^{-\phi(\l\Phi)}\sqrt{-{\rm det}(P[g_{ab}(\l\Phi)+
B_{ab}(\l\Phi)]+\l F_{ab}}\labell{biact}\,\,,
\eeqa
where we have defined $\l=2\pi\alpha'$, and the D-brane tension 
is $T_p=(g_s(2\pi)^p(\alpha')^{(p+1)/2})^{-1}$. Here, $F_{ab}$ is the 
abelian field
strength of the world-volume  gauge field, while
the metric and antisymmetric tensors are
the pull-backs of the bulk tensors to the D-brane world-volume, \eg
\beqa
P[g_{ab}]&=&
g_{ab}+2\l g_{i(a}\,\prt_{b)}\Phi^i+\l^2g_{ij}\prt_a\Phi^i\prt_b\Phi^j\,\,,
\labell{pull}
\eeqa
where  we have used that fact that we are employing static
gauge throughout the paper, \ie $X^a(x)=x^a$ for world-volume and 
$\l\Phi^i(x^a)$ for
transverse coordinates. The action \reef{biact} describes the 
dynamics of the open string fields, \ie $A_a$ and $\Phi^i$, and their 
couplings with each other and with the closed string fields. Dynamics of 
the closed string fields, on the other hand,  are described by the bulk 
supergravity action in which we are not interested in this paper. The 
derivatives of the gauge field strength, and the second derivatives of 
the scalars  are not included in this action. However, 
 the action includes the transverse
 derivatives of the closed string fields 
through the Taylor expansion of these fields 
\cite{hull,gm}, \eg
\beqa
\phi(\l\Phi)&=&\phi^0+\l\Phi^i\prt_i \phi^0+
\frac{\l^2}{2}\Phi^i\Phi^j\prt_i\prt_j \phi^0+
\frac{\l^3}{3!}\Phi^i\Phi^j\Phi^k\prt_i\prt_j\prt_k \phi^0+\cdots\,\,
\nonumber
\eeqa 
where the subscript $0$ means that the closed string field should be evaluated 
at the position of the brane, \ie $x^i=0$. 
Appearance of the above derivative terms in the DBI action 
verified in \cite{gm} by explicit evaluation of string theory S-matrix 
elements. At the same time the string theory calculation confirms that 
there is no analogous  world-volume derivative terms  in the action.

On the other hand, the coordinates of the D$_p$-brane with background 
constant B-flux becomes non-commutative\cite{acmrd,mrd,faha,csc,sw}. 
The leading order effective action of the D$_p$-brane in the SW limit is 
the non-commutative $U(1)$ Yang-Mills theory. In this action the symmetric 
part of $(g+\l B)^{-1}$, where $g$ and $B$ are the background constant 
fields, plays 
the role of the world-volume metric and the antisymmetric part plays 
the role of the 
non-commutative parameters.  Higher order $\alpha'$ correction to this 
 action for the case that the derivative of the 
non-commutative gauge field strength is small is the Born-Infeld 
generalization of the non-commutative 
Yang-Mills theory\cite{sw}. In this approximation,
however,  
one should use the ordinary product as the multiplication rule between 
fields in the Born-Infeld action. Extension of this Born-Infeld action to 
the action which includes
 the couplings of the non-constant closed string fields to the D$_p$-brane is 
not trivial, see \cite{hl2} for a proposal.

The linear coupling of the graviton to the non-commutative gauge fields and 
scalar fields of D$_p$-brane was 
found in \cite{yh1} by explicit evaluation of the disk S-matrix elements of 
one graviton and infinite number of massless open string states in the SW 
limit. This calculation shows that the   couplings  
in the superstring theory 
and  in the  
bosonic theory  are not identical. 
We shall find in the present  paper
 a non-commutative action which is  the transformation of 
the DBI  action \reef{biact} under the Seiberg-Witten 
 map with one open Wilson contour 
(see eq.\reef{sone}). This action  includes 
linear and  non-linear couplings of the massless closed string fields to 
the D-branes of the superstring theory.  The linear couplings in the SW limit 
are  exactly those that were found in \cite{yh1}. 

Our strategy to find the above mentioned action is that we start with 
expanding the action \reef{biact} which properly includes the non-constant 
 closed 
string fields. 
 We then 
transform the open string commutative fields in this action to the 
non-commutative fields using the  SW map. 
Since the SW map transforms each commutative open string  field in 
the DBI action \reef{biact}  to  
infinite
number of non-commutative  fields that  smear along  one
open Wilson contour\cite{liu,yh3,sm1,hl1}, the resulting action has 
infinite number of  open Wilson contours. 
On the other hand, we know that for  evaluating the string theory 
 S-matrix elements corresponding to the disk level effective action,
 the open string vertex operators  must be inserted  on the 
boundary of the disk world-sheet. Moreover, from the results 
in \cite{tmmw,mg1}, we  know that  the boundary of the disk world-sheet 
corresponds to one open Wilson contour. Hence, to map the disk level 
effective action \reef{biact} to another disk level action, we impose 
an extra operation that reduces the infinite number of 
Wilson lines  
 to only one  line. 
The first attempt 
in this direction was made in \cite{mg1}. 
In \cite{mg1} 
we found that in the transformed action the closed string fields should 
be functional of the non-commutative gauge fields as well as the 
non-commutative scalar fields. 
This feature was then confirmed by explicit 
evaluation of disk $S$-matrix elements of one closed string and infinite 
number of open string states for the special value of the non-commutative 
parameters, \ie finite $\alpha'$ and very large B-flux. It was hard to prove 
that the action found in \cite{mg1}  is  invariant under the non-commutative 
gauge transformation. We shall show in this paper, using the result 
in \cite{liu,sd}, that  the transformed action can be written in a manifestly 
gauge invariant form.

Although the action \reef{biact}  is valid only when the derivative  
of $F_{ab}$ is small, in the transformed action there is no such a 
limitation on $\hF_{ab}$, the non-commutative gauge field strength.  
The transformed action includes derivatives of $\hF_{ab}$ through
the $*$-product between $\hF$'s, couplings 
of the massless non-constant 
closed string fields to the $\hF$'s, a Wilson line, and a Wilson 
 operator. Interestingly, in the SW limit the new action describes 
properly the non-commutative super Yang-Mills theory. The part of the action 
which includes only the open string fields reduces to the non-commutative 
super YM action. And  
 the linear coupling of the graviton to the D-brane reduces 
exactly to the energy-momentum tensor of the non-commutative super YM 
theory found in \cite{yh1,yh2}. 
In \cite{yh1,yh2}, the energy-momentum tensor was calculated in string theory
by explicit evaluation of the disk S-matrix element of one graviton
and infinite number of open string gauge fields in the SW limit. 
In this calculation there is no limitation on $\hF_{ab}$, and the 
derivatives of $\hF_{ab}$ appear only through the $*$-product and 
the Wilson line. This indicates that the proposed non-commutative 
action has no further derivative correction terms in the SW limit.
 
The reminder of the paper is organized  as follows:  We begin in section 2 
by reviewing the construction of the non-commutative D$_9$-brane action 
proposed  in \cite{mg1}. We then construct the action for D$_p$-brane 
using the T-duality transformation rules. In section 3, using the 
important result in \cite{liu,sd}, we write the action in a manifestly 
non-commutative gauge invariant form. We extract the linear couplings of the
graviton to the non-commutative gauge fields for arbitrary non-commutative 
parameters in section 4. In section 5, we take the SW limit of the graviton 
couplings and show that they are reduced to the energy-momentum tensor of 
the non-commutative super YM theory found in \cite{yh1,yh2}. 
 In section 5.1, we write  the non-commutative super YM theory  
 in terms of the 
commutative fields, and compare it with the theory in terms of 
the non-commutative 
fields. 
We conclude with a brief discussion 
of our results in section 6.

\section{Transformation of  DBI action under the SW map 
with one  Wilson line}

In \cite{mg1}, we found an expression for transforming commutative DBI 
action of D$_p$-brane \reef{biact} under SW map with one open Wilson 
contour for any $p$. In that paper 
we did not check the consistency of the proposed action with T-duality and 
hence the 
commutators of two non-commutative scalar fields are  not included in 
that action. To include properly these terms  into the action, we start with 
the proposed non-commutative action for the $D_9$-brane which has no scalar 
field. We then transform it to the action for $D_p$-brane using the 
T-duality transformation rules. So we begin with a brief review of 
the construction of the non-commutative $D_9$-brane as follows:
\\
1. Start with the commutative $D_9$-brane action in which the 
non-constant closed string fields are included, that is,
\beqa
S&=&-T_9 \int d^{10}x\, e^{-\phi}\sqrt{-{\rm det}(g_{\mu\nu}+
B_{\mu\nu}+\l F_{\mu\nu})}\labell{biactone}\,\,.
\eeqa
\\
2. Expand the above action for non-constant quantum fluctuations around the 
constant background fields\footnote{Note that we have 
used the same symbols for the 
constant background fields and  for the the whole closed string fields.} 
$g_{\mu\nu}+\l B_{\mu\nu}$. For example,  the expansion 
for the linear dilaton is
\beqa
\cL(\phi,A)&=&T_9 \l c\,\phi\left(\frac{1}{2}\Tr(VF)
\right)\,\,,\nonumber\\
\cL(\phi,2A)&=&T_9 \l^2c\, \phi\left(-\frac{1}{4}\Tr(VFVF)+
\frac{1}{8}(\Tr(VF))^2)\right)\,\,,
\nonumber
\\
\cL(\phi,3A)&=&T_9\l^3 c\,\phi\left(\frac{1}{6}\Tr(VFVFVF)-
\frac{1}{8}\Tr(VF)\Tr(VFVF)+\frac{1}{48}(\Tr(VF))^3\right)\,\,,\nonumber
\eeqa
where the constants $c$ and $V^{\mu\nu}$-matrix are  defined as
\beqa
c\equiv \sqrt{-\det(g_{\mu\nu}+\l B_{\mu\nu})}\,\,\,\,\,&;&\,\,\,\,\,
V^{\mu\nu}\equiv \left(\frac{1}{g+\l B}\right)^{\mu\nu}\,\,.
\nonumber
\eeqa
\\
3. Transform each commutative gauge field strength $F_{ab}$ to 
non-commutative gauge field $\hA_a$ and $\hF_{ab}$ according to 
the SW map\cite{mine,tmmw}
\beqa
F_{\mu\nu}&=&\hF_{\mu\nu}+\theta^{\alpha\beta}
\left(\prt_{\beta}(\hA_{\alpha}\hF_{\mu\nu})-
\frac{1}{2}\hF_{\mu\nu}\hF_{\alpha\beta}-
\hF_{\mu\alpha}\hF_{\nu\beta}\right)*_2\labell{trans3}\\
&&+\frac{1}{2}\theta^{\alpha\beta}\theta^{\gamma\delta}
\left(\prt_{\alpha}\prt_{\gamma}(\hF_{\mu\nu}\hA_{\beta}\hA_{\delta})-
\prt_{\gamma}(\hF_{\alpha\beta}\hF_{\mu\nu}\hA_{\delta})+
2\prt_{\gamma}(\hF_{\mu\alpha}\hF_{\nu\beta}\hA_{\delta})\right.\nonumber\\
&&\left.-(\hF_{\mu\alpha}\hF_{\nu\beta}\hF_{\gamma\delta})+
\frac{1}{4}(\hF_{\mu\nu}\hF_{\alpha\beta}\hF_{\gamma\delta})+
\frac{1}{2}(\hF_{\mu\nu}\hF_{\beta\gamma}\hF_{\alpha\delta})-
2(\hF_{\alpha\gamma}\hF_{\mu\delta}\hF_{\nu\beta})\right)*_3+O(\hA^4)\,\,.
\nonumber
\eeqa
Appearance of the $*_N$ above shows that the non-commutative fields on 
the left hand side are smeared  along a  Wilson line. In the above 
transformation, $\hA_{\mu}$ is 
the non-commutative gauge field, and 
\beqa
\hF_{\mu\nu}&=&\prt_{\mu}\hA_{\nu}-\prt_{\nu}\hA_{\mu}-i\hA_{\mu}*\hA_{\nu}+
i\hA_{\nu}*\hA_{\mu}\nonumber\\
&=&\prt_{\mu}\hA_{\nu}-\prt_{\nu}\hA_{\mu}-i[\hA_{\mu},\hA_{\nu}]_M\,\,,
\nonumber
\eeqa
is the non-commutative gauge field strength. The $*$-product  is 
\beqa
f(x)*g(x)&=&e^{\frac{i}{2}\theta^{ab}\frac{\prt}{\prt x^a_1}
\frac{\prt}{\prt x^b_2}}f(x_1)g(x_2)|_{x_1=x_2=x}\,\,.
\nonumber
\eeqa
\\
4. After transforming all the commutative fields according to the above 
SW map, one finds infinite number of   Wilson lines. We perform an 
extra operation that reduces the infinite number of  Wilson lines  
to only one line, \ie  insert  the $*_N$ product 
between N non-commutative 
gauge fields and $\hF$'s. This stems from the fact that the disk 
world-sheet has only one boundary. After some manipulations, and benefit of 
the following identity\cite{liu}
\beqa
\theta^{\mu\nu}\prt_{\mu}\left(f_1...f_N\prt_{\nu}g\right)*_N&=&i
\sum_{j=1}^{N-1}\left(f_1...[f_j,g]_M...f_{N-1}\right)*_{N-1}\,\,,
\labell{idenone}
\eeqa
one finds
\beqa
{\hat{\cL}}(\phi,\hA)&=&T_{9} c\,\prt_{\mu}\phi(\hA^{\mu})\,\,,\nonumber\\
{\hat{\cL}}(\phi,2\hA)&=&T_9 c\,\left(\frac{1}{2}\prt_{\mu}\prt_{\nu}
\phi(\hA^{\mu}\hA^{\nu})*_2-\frac{\l^2}{4}\phi\Tr(G\hF G\hF)*_2\right)\,\,,
\nonumber\\
{\hat{\cL}}(\phi,3\hA)&=&T_9 c\,\left(\frac{1}{3!}\prt_{\mu}\prt_{\nu}
\prt_{\alpha}\phi(\hA^{\mu}\hA^{\nu}\hA^{\alpha})*_3-\frac{\l^2}{4}
\prt_{\mu} \phi\left(\hA^{\mu}\Tr(G\hF G\hF)\right)*_3\right)\,\,,
\nonumber
\eeqa
where we have defined $\hA^{\mu}=\theta^{\mu\nu}\hA_{\nu}$, and 
\[
\theta^{\mu\nu}=\l\left(\frac{1}{g+\l B}\right)^{\mu\nu}_A\,\,\,\,\,;\,\,\,\,\,
G^{\mu\nu}=\left(\frac{1}{g+\l B}\right)^{\mu\nu}_S\,\,,
\]
and $()_A$ and $()_S$ denote the antisymmetric and symmetric part of 
the $V$-matrix.  In writing the  terms in the above equations  we have 
ignored some total derivative terms, \eg we have written 
$(1/2)\prt_{\mu}\prt_{\nu}\phi(\hA^{\mu}\hA^{\nu})*_2$ instead of 
$(1/2)\phi\prt_{\mu}\prt_{\nu}(\hA^{\mu}\hA^{\nu})*_2$.

Note that  in the above Lagrangians only 
the symmetric part of the $V$-matrix  plays the role of 
metric for the open string field $\hF_{\mu\nu}$. 
This is  consistent with the fact that the $G$
 is the open string metric\cite{sw}.
The terms that involve the antisymmetric part of the $V$-matrix, \ie  
$\theta/\l$,  combine in such a way that they make  the closed string 
field to be  functional of the non-commutative gauge field, \ie functional 
of $\theta^{\mu\nu}\hA_{\nu}$. If one does similar calculation for graviton, 
one again  finds that  $G^{\mu\nu}$ appears only when both indices of 
the $V$-matrix contract with $\hF$'s, otherwise the matrix $V^{\mu\nu}$ 
plays the role of rising the indices.  At the same time the graviton, like 
the dilaton, becomes the   functional of the non-commutative gauge field.   
All the resulting terms can be reproduced by  the following prescribed 
action:
\beqa
\hS&=&-T_{9} \int d^{10}x
\left(e^{-\phi(\hA)}\sqrt{-\det\left(g_{\mu\nu}(\hA)+
B_{\mu\nu}(\hA)+\l\hF_{\mu\nu}\right)}\right)*_N\,\,.
\labell{nonbiact}
\eeqa
Expanding above  action around the constant background fields 
for non-constant quantum 
fluctuations, one finds  various couplings between open and closed 
string fields that  matrix $V=(g+\l B)^{-1}$ plays the role of metric. 
The prescription is that when both indices of $V^{\mu\nu}$ contract with the 
open string  fields, one must replace it with the symmetric part of the 
$V$-matrix\footnote{Needless to mention that when both indices of $V^{\mu\nu}$ 
contract with one $\hF$ the result is zero.}, that is
\beq
V^{\mu\nu}\longrightarrow G^{\mu\nu}\,\,\, 
{\rm {when \,\,both \,\,indices\,\,contract\,\,with \,\, 
the\,\,open\,\,string\,\,fields.}}
\labell{vtog}
\eeq
The resulting terms are then fully consistent with the terms in part 4. 
above.
  When there is no background B-flux, the above non-commutative action 
reduces to the ordinary BI action \reef{biactone} with no B-flux.  
When there is no non-constant
closed string field, only the symmetric part of the $V$-matrix appears 
in the expansion, and  the $*_N$ product reduces to $*$ product\cite{mg1}. 
Consequently  the action \reef{nonbiact} is consistent with  
  the non-commutative BI action found  
in \cite{sw}. Hence it was conjectured in \cite{mg1} that  the  
action \reef{nonbiact} is the correct transformation of the ordinary BI 
action \reef{biactone} under the SW map with one Wilson line.

To extend this action to the action for $D_p$-brane, we use the familiar 
rules of T-duality. 
 We now assume that the B-flux is non-zero only in $0,1,\cdots p$ 
directions. 
 To apply T-duality we also assume that all fields are independent 
of $p+1,\cdots, 9$ directions. Since the non-commutative  parameter 
$\theta^{ab}$  that appears in the definition of $*_N$ product 
(see eq.\reef{starn}) and in the functional dependence of the closed 
strings to $\hA^a$,   have  no component in the $p+1,\cdots, 9$ directions, 
these parts of action do not  change under T-duality transformation.  The 
transformation of the  other parts of the action \reef{nonbiact} under 
T-duality  are  exactly like the non-abelian commutative cases that were  
studied in details by Myers in \cite{rm1}. The only change to that analysis 
is that every non-abelian commutator must be replaced by Moyal commutator. 
Therefore, using the result of \cite{rm1}, one will find the following T-dual 
action: 
\beqa
\hS&=&-T_{p}\int d^{p+1}x\left(e^{-\phi(\hA)}\right.\nonumber\\
&&\left.\times \sqrt{-\det\left( P_{\theta}[E_{ab}(\hA)+E_{ai}(\hA)(Q^{-1}-
\delta)^{ij}E_{jb}(\hA)]+\l\hF_{ab}\right)\det (Q^i{}_j)}\right)*_N\,\,,
\labell{lllone}
\eeqa
where the closed string field $E_{\mu\nu}=g_{\mu\nu}+B_{\mu\nu}$, the 
transverse indices $i,j,\cdots$ are raised by the inverse of $E_{ij}$, 
\ie $E^{ij}$  and 
\[
Q^i{}_j=\delta^i{}_j-i\l[\hP^i,\hP^k]_ME_{kj}(\hA)\,\,.
\]
The definition of the pull-back $P_{\theta}$ is the extension 
of \reef{pull} in which ordinary derivative is replaced by its 
non-commutative covariant derivative, \ie 
$\prt_a\Phi^i\rightarrow D_a\hP^i=\prt_a\hP^i-i[\hA_a,\hP^i]_M$.  
In the action \reef{lllone}  one must also use the prescription 
\reef{vtog}.

Now that we have found   the T-dual action, similar to the non-abelian 
commutative case, we assume that the closed string fields depend on the 
$p+1,\cdots,9$ directions, and  further generalize it by assuming that the 
closed string fields are functional of the non-commutative scalar fields 
$\hP^i$ \cite{gm}, \ie,  $\phi(\hA)\rightarrow \phi(\hA,\l\hP)$ and 
$E_{\mu,\nu}(\hA)\rightarrow E_{\mu\nu}(\hA,\l\hP)$. The functional dependence 
of the  closed string fields on  both $\hA^a$ and $\hP^i$ was also found 
in \cite{mg1} by direct evaluation of the disk S-matrix elements, and by 
transforming the ordinary DBI action under the SW map with one Wilson line.  

The action \reef{lllone} may be extended to the non-abelian case by 
converting the open string fields to the matrix valued fields. Since the 
covariant derivative of the transverse scalar fields and commutator of two 
scalar fields are 
already included  in the action in such a way that they are 
consistent with T-duality transformation rules, one needs only to define a 
prescribed trace over the matrices. The prescription for the trace should 
be in such a way that when  $B=0$ the result  reduces to the non-abelian 
commutative case. In that case the prescription for the trace is the 
symmetrized trace \cite{aat,rm1}. In particular, after expanding the 
action for quantum fluctuations,  one should write each terms 
of the expanded 
action in a form that is symmetric between $F_{ab}$, $D_a\Phi^i$, 
$[\Phi^i,\Phi^j]$ and individual $\Phi^i$. The later field coming from Taylor 
expansion of the closed string fields. Then take the trace of the resulting 
terms. Hence, in the 
non-abelian non-commutative case, our prescription is the 
symmetrized trace over $\hF_{ab}$, $D_a\hP^i$, $[\hP^i,\hP^j]_M$ which
are coming from expanding the square root in \reef{lllone}, and 
individual $\hA^a$, $\hP^i$ which are coming from the Taylor expansion of the 
closed string fields. The resulting action reduces to the commutative case 
when $B=0$,  because in this  case $*_N$-product $\longrightarrow $ 
ordinary product, 
$\hA^a=\theta^{ab}\hA_b\longrightarrow 0$,  and 
$V^{ab}\longrightarrow G^{ab}$. 

The above symmetrized trace prescription make sense only  when the 
multiplication rules between $\hF_{ab}$, $D_a\hP^i$, $[\hP^i,\hP^j]_M$,  
$\hA^a$, and $\hP^i$ are   symmetric under permutation of these fields. 
In fact it was shown in \cite{liu} that  the $*_N$ satisfies this condition. 
Therefore in the non-abelian extension of the 
action \reef{lllone}, the $*_N$ must be the multiplication 
rule between fields $\hF_{ab}$, $D_a\hP^i$, $[\hP^i,\hP^j]_M$, $\hA_a$ 
and $\hP^i$. This is also consistent with the step 4 in our construction 
of the non-commutative D$_9$-brane action.
 
\section{Action in gauge invariant form }

The transformed DBI action under the SW map with one Wilson line
in the form appearing in 
\reef{lllone} is not manifestly invariant under the non-commutative gauge 
transformation. In this section we would like to 
 write it  in a manifestly gauge invariant form. To this end, we will use 
the following observation \cite{liu,sd}. Given a collection of open string 
local operators $\hQ_i(x^a)$ on the world-volume of the $D_p$-brane which 
transform in the adjoint representation of non-commutative $U(1)$ gauge 
transformation, one can obtain a natural gauge invariant operator of fixed 
momentum $k^a$ by smearing these local operators along the  straight contour 
$\z^a(\tau)=(\theta^{ab}k_b)\tau$ with $0\leq\tau\leq 1$, and multiplying 
the product by a Wilson  
operator
 along the same contour. Following this observation, we consider the 
 Wilson  operator
\beqa
W(k^{\mu},x^a,C)&=&e^{i\int_0^1d\tau\left(k_a\theta^{ab} \hA_a(x+\z(\tau))+
\l k_i\hP^i(x+\z(\tau))\right)}\,\,,
\labell{wwone}
\eeqa
and the following  collection of local operators  $\hQ_i(x^a)$ that transform 
in the adjoint and local closed string fields $O_j(x^{\mu})$ that are 
scalar under the non-commutative gauge transformation:
\beqa 
f(x^{\mu})&=&\prod_{j=1}^m O_j(x^{\mu})
\prod_{i=1}^{n}\hQ_i(x^a)\,\,.
\nonumber
\eeqa
According to the result in \cite{liu,sd}, the following operator is gauge 
invariant
\beqa
\tf_W(x^a,k^{\mu})&=&\int d^{10}y\,
\prod_{j=1}^m O_j(x^a,y^i)
\labell{wfive}\\
&&\times P_*\left[ W(k^{\mu},y^a,C)
\prod_{i=1}^n\int_0^1d\tau_i\hQ_i(y^a+\z^a(\tau_i))\right]*e^{ik\cdot y}
\nonumber\\
&=&\int d^{10}y\,
\prod_{j=1}^m O_j(x^a,y^i)
L_*\left[ W(k^{\mu},y^a,C)
\prod_{i=1}^n\hQ_i(y^a)\right]*e^{ik\cdot y}\,\,,\nonumber
\eeqa
where $k\cdot y=k_{\mu}y^{\mu}$ and $P_*$ denotes path-ordering with respect 
to the $*$-product, while $L_*$ is an abbreviation for the combined 
path-ordering and integrations over $\tau$'s. In this formula the open 
string operators $\hQ_i$ are smeared over the straight contour of the Wilson 
line while the locations of  the closed string operators $O_j$ are independent 
of  the contour. Note that the $y$-integral is over the world-volume 
position of the open string operators and over the transverse position of 
the closed string fields. There is no integral over
 the world-volume position of the 
closed string fields.

Now  we expand the exponential in the Wilson operator,
 and then perform the line 
integrals over  $\tau$'s. Following \cite{liu}, one may Fourier transforms  
the open string operators 
\beqa
\cO_i(y^a+\z^a(\tau_i))&=&\int\frac{d^{p+1}k_i}{(2\pi)^{p+1}}
\tcO_i(k_i)e^{-i(k_i)_a(y^a+\z^a(\tau_i))}\nonumber\\
&=&\int\frac{d^{p+1}k_i}{(2\pi)^{p+1}}\tcO_i(k_i)e^{-i(k_i)_a 
y^a-i(k_i\times k)\tau_i}\,\,,
\nonumber
\eeqa
where $k_i\times k=(k_i)_a\theta^{ab}k_b$, and $\cO_i$ is any of $\hP$'s, 
$\hA$'s or $\hQ$'s. The integral over $\tau$'s converts to an elementary 
exponential integral which yields the $*_N$-product  in the momentum 
space\cite{hljm,liu}, that is\footnote{An alternative formula for the 
$*_N$-product was found in \cite{mg1}.}
\beqa
\left(\tcO_1\cdots \tcO_N\right)*_N&=&
\int_{0}^{1} d\tau_1\int_{0}^{1}d\tau_2\cdots\int_{0}^{1} d
\tau_N \labell{starn}\\
&&\exp\left(-i\sum_{i=1}^N(k_i\times k)\tau_i+\frac{i}{2}
\sum_{i<j}^{N}(k_i\times k_j)\epsilon(\tau_{ij})\right)
(\tcO_1\cdots \tcO_N)\,\,,
\nonumber
\eeqa
where $\epsilon(\tau_{ij})=+1(-1)$ for $\tau_{ij}>0$($\tau_{ij}<0$), and 
$\tau_{ij}=\tau_i-\tau_j$. Note that the integral over $y^a$ yields the 
condition $k^a=\sum_i (k_i)^a$. Replacing all the momenta in the expansion 
with their appropriate derivatives and then using 
\beqa
\int\frac{d^{p+1}k_i}{(2\pi)^{p+1}}\tcO_i(k_i)e^{-i(k_i)_a y^a}&=&
\cO_i(y^a)\,\,,
\nonumber
\eeqa
one will find
\beqa
\tf_W(x^a,k^{\mu})&=&\int d^{10}y\,\,e^{ik\cdot y}
\sum_{p=0,q=0}^{\infty}\frac{\l^p}{p!q!}(\prt_{y^{i_1}}
\cdots\prt_{y^{i_p}})\left(\prod_{j=1}^mO_j(x^a,y^i)\right)\nonumber\\
&&\times(\prt_{y^{a_1}}\cdots\prt_{y^{a_q}})\left( \hP^{i_1}
\cdots\hP^{i_p}\hA^{a_1}\cdots\hA^{a_q}\hQ_1\cdots \hQ_n\right)*_{p+q+n}
\,\,,
\nonumber
\eeqa
where now the $*_N$ is in the position space 
( recall that  $\hA^a=\theta^{ab}\hA_b$).  
The  
Fourier  inverse  of the above function is
\beqa
f_W(x^{\mu})&=&\int \frac{d^{10}k}{(2\pi)^{10}} 
\tf_W(x^a,k^{\mu})e^{-ik\cdot x}\nonumber\\
&=&\sum_{p=0,q=0}^{\infty}\frac{\l^p}{p!q!}(\prt_{x^{i_1}}\cdots
\prt_{x^{i_p}})\left(\prod_{j=1}^mO_j(x^a,x^i)\right)\nonumber\\
&&\times(\prt_{x^{a_1}}\cdots\prt_{x^{a_q}})\left( \hP^{i_1}
\cdots\hP^{i_p}\hA^{a_1}\cdots\hA^{a_q}\hQ_1\cdots \hQ_n\right)*_{p+q+n}
\nonumber\\
&=&\left(\prod_{j=1}^mO_j(\hA,\l\hP)\prod_{i=1}^n\hQ_i\right)*_N+
({\rm total\,\,world\,\,volume\,\,derivative\,\,terms})\,\,.
\nonumber
\eeqa
The total world-volume derivative terms above are exactly like the total 
derivative terms that we ignored  in the step 4 in the section 2 for  
constructing the non-commutative action. The world-volume 
integral of $f_W$ that we are interested in is
\beqa
\hS'&\equiv&\int d^{10}x\, f_W(x^{\mu})\delta(x^i)\nonumber\\
&=&\int d^{p+1}x\, \left(\prod_{j=1}^mO_j(\hA,\l\hP)
\prod_{i=1}^n\hQ_i\right)*_N\,\,.
\labell{weight}
\eeqa
Therefore, expanding the Wilson operator and performing the Wilson 
line integral, one finds that the closed string fields become functional
of the non-commutative scalar and gauge fields, and the $*_N$ operates
as the multiplication rule between the open string fields.

On the other hand,  if one does not perform the $\tau$ integral in 
\reef{wfive}, one will find that the Wilson line and operator 
appear in  the action, 
that is
\beqa
\hS'&=&\int d^{p+1}x \frac{d^{10}k}{(2\pi)^{10}} d^{10}y\,e^{-ik_a x^a}
\prod_{j=1}^mO_j(x^a,y^i)\nonumber\\
&& \times L_*\left[W(k^{\mu},y^a,C)\prod_{i=1}^n\hQ_i(y^a)
\right]*e^{ik\cdot y}\,\,.
\labell{Stwo}
\eeqa
One may wish to  write the closed string fields in the 
above action in the momentum space, that is 
\beqa
O_j(x^a,y^i)&=&\int \frac{d^{10}p_j}{(2\pi)^{10}}
\tO_j(p_j^{\mu})e^{-i(p_j)_a x^a-i(p_j)_i y^i}\,\,.
\nonumber
\eeqa
In this case, one can  perform the integral over $x^{a}$,  $y^i$ 
and then over $k^{\mu}$. 
The result simplifies to
\beqa
\hS'&=&\prod_{j=1}^m\int\frac{d^{10}p_j}{(2\pi)^{10}}
\tO_j(p_i^{\mu})\int d^{p+1}y\,L_*\left[W(k^{\mu},y^a,C)
\prod_{i=1}^n\hQ_i(y^a)\right]*e^{ik_ay^a}\,\,,
\nonumber
\eeqa
where now the components of the momentum of the Wilson operator 
is $k^{a}=-\sum_j (p_j)^{a}$ and $k^i=\sum_j (p_j)^i$. From the 
$y^a$-integral, on the other hand,  we  find  that the Wilson line 
momentum $k^a=\sum_i (k_i)^a$. Hence, there is the  momentum conservation in 
the world-volume directions, \ie $\sum_i(k_i)^a+\sum_j(p_j)^a=0$, whereas 
there is no momentum conservation in the transverse direction. This stems 
from the fact that we fixed the position of the $D_p$-brane at $x^i=0$.

The  structure in \reef{weight} is exactly the one  appears in the proposed 
non-commutative action in \reef{lllone},  when one expands the square root 
in the action. Hence, an alternative way of writing the non-commutative 
action \reef{lllone} is 
\beqa
\hS&=&\int d^{p+1}x \frac{d^{10}k}{(2\pi)^{10}}
 d^{10}y\,e^{-ik_a x^a}L_*\left[W(k^{\mu}, y^a,C)\hL(x^a,y^{\mu})\right]*
e^{ik\cdot y}\,\,,
\labell{sone}
\eeqa
where
\beqa
\hL(x^a, y^{\mu})&=&
-T_{p}\,e^{-\phi}
\sqrt{-\det\left( P_{\theta}[E_{ab}+E_{ai}(Q^{-1}-\delta)^{ij}E_{jb}]+
\l\hF_{ab}\right)\det (Q^i{}_j)}\,\,,
\labell{llllone}
\eeqa
and
\[
Q^i{}_j=\delta^i{}_j-i\l[\hP^i,\hP^k]_ME_{kj}\,\,.
\]
The closed string fields in the above action 
are function of $x^a,y^i$, whereas, 
the open string fields  are function of $y^a$. In \reef{llllone}, 
one must also take the prescription \reef{vtog} into  
account. When expanding the square root in \reef{llllone}, one finds a 
tower of different terms. Each term may contain 
 the open string operators $\hF_{ab}$, $D_a\hP^i$ or $[\hP^i,\hP^j]_M$, 
as well as massless closed string fields.   The position of the closed 
string fields are independent of the Wilson line, and the multiplication 
rule between them  is also the ordinary product. Hence, the $L_*$ has no 
effect on the closed string fields in each term of the expansion, and its 
effect on the open string operators is that it smears them along the open 
Wilson line. 
 Since the open string operators $\hF_{ab}$, $D_a\hP^i$ and $[\hP^i,\hP^j]_M$ 
transform in the adjoint under the gauge transformation, the action 
\reef{sone}, like \reef{Stwo},  is manifestly non-commutative gauge invariant.

Extension of this form of action \reef{sone} to the non-abelian case is 
again straightforward. We need to change the fields to matrix valued 
fields and a prescribed trace 
over the matrices. The prescription is again the symmetrized trace over 
$\hF_{ab}$, $D_a\hP^i$, $[\hP^i,\hP^j]_M$ which are coming from the expansion 
of the square root in \reef{llllone}, and $\hA^a$, $\hP^i$ which are coming 
from the expansion of the Wilson operator in \reef{sone}. In this case, 
however, the symmetrized trace is reproduced by the ordinary trace and 
the path ordering prescription. 
The path ordering means that after expanding the square root and the 
Wilson operator, one must integrate all different sequences of 
the open string operators  along the Wilson line. Taking the trace of 
the resulting terms produces the symmetrized trace. 

\section{Linear  Couplings}

The action \reef{sone} includes all linear and non-linear couplings of 
the closed string fields to the non-commutative D$_p$-branes. We would 
like to find the linear couplings of the dilaton and the graviton to the 
branes. So we write the closed string fields as the classical constant
background 
fields plus their non-constant 
quantum fluctuations, \ie $g_{ab}=g_{ab}+2\ka h_{ab}$, 
$g_{ai}=2\ka h_{ai}$, $g_{ij}=g_{ij}+2\ka h_{ij}$, $\phi=2\ka\phi'$, 
$B_{ab}=\l B_{ab}$, and the other components of the B-flux are 
zero. 
Now we expand \reef{llllone} around the background fields $g_{ab}+\l B_{ab}$. 
For the dilaton one finds,
\beqa
\hL_{\phi}&=&2T_p\ka c\,\phi'
\sqrt{\det\left(\delta^a{}_b+\l^2(Q^{-1})_{ij}G^{ac}D_c\hP^i D_b\hP^j+
\l G^{ac}\hF_{cb}\right)\det(Q^i{}_j)}\,\,,
\labell{lzero}
\eeqa
where
\[
Q^i{}_j=\delta^i{}_j-i\l[\hP^i,\hP_j]_M\,\,,
\]
where we have factored out the constant $c=\sqrt{-\det(g_{ab}+\l B_{ab})}$ 
from the square root above, and used the prescription \reef{vtog}. The square 
root involves only the open string fields in this case, hence, one must 
replace the $V^{ab}$ matrix by the open string metric $G^{ab}$ in all terms. 
In the above and subsequent  equations the indices  of the transverse scalar 
fields are lowered by the background field $g_{ij}$. 

For the graviton, because of the prescription \reef{vtog}, it is hard to find 
a closed form for its linear coupling to the D-brane. So we expand the square 
root in \reef{llllone} and keep terms that involve linearly the graviton. 
The expansion is straightforward, and the result is,
\beqa
\hL_h\!\!\!&=&\!\!\!-T_p\ka c\left(V^{ab}h_{ba}\left(1+\frac{\l^2}{2}
G^{cd}D_c\hP_i D_d\hP^i-\frac{\l^2}{4}G^{cd}\hF_{de}G^{ef}\hF_{fc}+
\frac{\l^2}{4}[\hP^i,\hP^j]_M[\hP_j,\hP_i]_M\right)\right.\nonumber\\
&&- V^{ab}h_{bc}V^{cd}\left(\l \hF_{da}+\l^2 D_d\hP_i D_a\hP^i-\l^2
\hF_{de}G^{ef}\hF_{fa}\right)\labell{lthree}\\
&&+\l^2V^{ab}h_{ij}D_a\hP^i D_b\hP^j+2\l V^{ab}h_{i(b}D_{a)}\hP^i-
2\l^2V^{ab}h_{i(b}D_{c)}\hP^i V^{cd}\hF_{da}\nonumber\\
&&\left.+2i\l^2 V^{ab}h_{i[a}[\hP^i,\hP^j]_MD_{b]}\hP_j+
\l^2[\hP^i,\hP^k]_M[\hP_k,\hP^j]_Mh_{ij}+\cdots\right)\,\,,
\nonumber
\eeqa
we have used  again the prescription \reef{vtog}. In this case though,  
the closed string graviton as well as  the open string fields appear in the 
expansion, so one  should not replace the $V^{ab}$ by $G^{ab}$ for all terms 
in the expansion above.  
In  eq.\reef{lthree}, dots represent terms that would be of order $\l^3$ 
if the $V$-matrix and $g_{ij}$  were independent of $\l$.  
Writing $V^{ab}=G^{ab}+\theta^{ab}/\l$, one finds
\beqa
\hL_h\!\!\!&=&\!\!\!-T_p\ka c\left(G^{ab}h_{ba}\left(1+
\frac{\l^2}{2}G^{cd}D_c\hP_i D_d\hP^i-\frac{\l^2}{4}G^{cd}\hF_{de}G^{ef}
\hF_{fc}+\frac{\l^2}{4}[\hP^i,\hP^j]_M[\hP_j,\hP_i]_M\right)\right.\nonumber\\
&&\- 2G^{ab}h_{bc}\theta^{cd} \hF_{da}-\l^2 G^{ab}h_{bc}G^{cd}D_d\hP_i 
D_a\hP^i-\theta^{ab}h_{bc}\theta^{cd}D_d\hP_i D_a\hP^i\nonumber\\
&&+\l^2G^{ab}h_{bc}G^{cd}\hF_{de}G^{ef}\hF_{fa}+\theta^{ab}h_{bc}
\theta^{cd}\hF_{de}G^{ef}\hF_{fa}\labell{lfinal}\\
&&+\l^2G^{ab}h_{ij}D_a\hP^i D_b\hP^j+2\l G^{ab}h_{ib}D_{a}\hP^i-
2\l \theta^{ab}h_{ib}D_{c}\hP^i G^{cd}\hF_{da}\nonumber\\
&&\left.+2i\l \theta^{ab}h_{ia}[\hP^i,\hP^j]_MD_{b}\hP_j+
\l^2[\hP^i,\hP^k]_M[\hP_k,\hP^j]_Mh_{ij}+\cdots\right)\,\,.\nonumber
\eeqa
According to the prescription \reef{vtog}, each  term in the expansion has 
at most two $V$-matrix, \ie the indices of other $V^{ab}$ contract with open 
string fields so they must be replaced by  $G^{ab}$. More specificly, terms 
that involve $h_{ab}$, $h_{ia}$ and $h_{ij}$ has  two, one and no $V^{ab}$ 
matrix. Therefore, assuming that $G^{ab}$, $\theta^{ab}$ and $g_{ij}$ are 
independent of $\l$, the  dots above gives corrections of order $O(\l)$, 
$O(\l^2)$ and $O(\l^3)$ to the terms that involve $h_{ab}$, $h_{ia}$ and 
$h_{ij}$, respectively.

Now replacing \reef{lzero} and \reef{lfinal} to the action \reef{sone},
transforming 
the quantum closed string fields to the momentum space,  
 and 
performing the integral over $x^a$, $y^i$ and $k^{\mu}$, 
one will find
\beqa
\hS_{\phi}&=&\frac{2\ka}{g_{YM}^2\l^2}\int\frac{d^{10}p}{(2\pi)^{10}}
\sqrt{-\det G}\left(\hat{\phi}'(p)\hat{T}_{\phi}(p)\right)\,\,,\nonumber\\
\hS_h&=&-\frac{\ka}{g_{YM}^2\l^2}\int\frac{d^{10}p}{(2\pi)^{10}}
\sqrt{-\det G}
\left(\hat{h}_{\mu\nu}(p)\hat{T}_{h}^{\mu\nu}(p)\right)\,\,,
\labell{s1one}
\eeqa
where
\beqa
\hat{T}(p)&=&\int d^{p+1}y L_*\left[W(k^{\mu},y^a,C)T_{\phi}(y)\right]*
e^{ik_a y^a}\,\,,\nonumber\\
\hat{T}^{\mu\nu}(p)&=&\int d^{p+1}y L_*\left[W(k^{\mu},y^a,C)T_h^{\mu\nu}(y)
\right]*e^{ik_a y^a}\,\,,
\nonumber
\eeqa
and 
\beqa
T_{\phi}&=&
\sqrt{\det\left(\delta^a{}_b+\l^2(Q^{-1})_{ij}G^{ac}D_c\hP^i D_b\hP^j+
\l G^{ac}\hF_{cb}\right)\det(Q^i{}_j)}\,\,,\nonumber\\
T_h^{ab}&=&G^{ab}\left(1+\frac{\l^2}{2}G^{cd}D_c\hP_i D_d\hP^i-
\frac{\l^2}{4}G^{cd}\hF_{de}G^{ef}\hF_{fc}+
\frac{\l^2}{4}[\hP^i,\hP^j]_M[\hP_j,\hP_i]_M\right)\nonumber\\
&&- 2G^{ca}\theta^{bd} \hF_{dc}-\l^2 G^{ca}G^{bd}D_d\hP_i D_c\hP^i-
\theta^{ca}\theta^{bd}D_d\hP_i D_c\hP^i\nonumber\\
&&+\l^2G^{ca}G^{bd}\hF_{de}G^{ef}\hF_{fc}+\theta^{ca}\theta^{bd}
\hF_{de}G^{ef}\hF_{fc}+\cdots\,\,,
\labell{tone}
\\
T_h^{ai}&=&\l G^{ba}D_b\hP^i-\l \theta^{ba}D_c\hP^i G^{cd}\hF_{db}
+i\l\theta^{ab}[\hP^i,\hP^j]_MD_b\hP_j+\cdots\,\,,\nonumber\\
T_h^{ij}&=&\l^2 G^{ab}D_a\hP^i D_b\hP^j+\l^2[\hP^i,\hP^k]_M[\hP_k,\hP^j]_M+
\cdots\,\,.\nonumber
\eeqa
Components of the momentum of the  Wilson  operator  
are  $k^a=-p^a$ and $k^i=p^i$. 
In the equation \reef{s1one}
, using the definition of the effective non-commutative Yang-Mills 
coupling \cite{sw}, we have written
\beqa
T_p c&=&\left(\frac{\sqrt{-\det(g_{ab}+\l B_{ab})}}{
\sqrt{-\det G}g_s(2\pi)^p(\alpha')^{\frac{p+1}{2}}}\right)
\sqrt{-\det G}\nonumber\\
&=&\frac{1}{g_{YM}^2\l^2}\sqrt{-\det G}\,\,.
\nonumber
\eeqa

The parameters in the 
energy-momentum tensors in \reef{s1one} are arbitrary.
These tensors  are fully consistent with the disk S-matrix 
element of two massless open strings and one closed string state, and of two 
massless closed string states  in 
the superstring theory\cite{mrg,shyk,mine,hl2}, for the arbitrary parameters.
 However, we would like to 
compare our result with the disk S-matrix element of one closed string and 
infinite number of the massless open string states. This calculation has 
been done in \cite{yh1} for the special values of the  
parameters, \ie the SW limit.  

\section{The Seiberg-Witten  limit}

In the Seiberg-Witten limit $\l\sim \sqrt{\epsilon}\rightarrow 0$, 
$g_{ab}\sim\epsilon\rightarrow 0$ and all other  closed string background 
fields are finite.
If the background B-flux has only space-space components, then the 
limit is
$\l\rightarrow 0$ and  $G^{ab}$, $\theta^{ab}$, and $g_{ij}$ 
are finite.
In this limit, all the closed strings  and the  massive open string 
fields decouple from the D-brane, and the entire string dynamics is 
described by the non-commutative U(1) YM field 
theory\cite{sw}\footnote{ 
The minimal change to our notation to include the background B-flux with 
only space-space components is the following:
We still use $\theta^{ab}$ for the whole world-volume directions. 
However, 
when one of the indices of $\theta^{ab}$ takes value in the commutative 
directions of the world-volume, it must be replaced by zero.}. In this 
limit, when there is no non-constant closed string field, 
 the action \reef{sone}  reduces to   
\beqa
\hS&=&-\frac{1}{g_{YM}^2}\int d^{p+1}x \,\sqrt{-\det G}*\left( 
-\frac{1}{4}G^{cd}\hF_{de}G^{ef}\hF_{fc}\right.\nonumber\\
&&\left.+\frac{1}{2}G^{cd}D_c\hP_i D_d\hP^i
+\frac{1}{4}[\hP^i,\hP^j]_M[\hP_j,\hP_i]_M\right)\,\,,
\labell{stwo}
\eeqa
where we have used the fact that the momentum of the Wilson operator 
\reef{wwone} is summation of the closed string fields. No closed string 
field, no Wilson operator and no Wilson line (recall that the 
length of Wilson line is proportional to the 
momentum of the Wilson operator).  This action  is  the non-commutative action 
introduced in \cite{sw}.

The linear coupling of the non-commutative YM theory to graviton and to 
the dilaton can be read from the general result in \reef{s1one}--\reef{tone} 
by taking the SW limit. It was argued in \cite{yh1} that  
the on-shell condition of the closed string fields, in the SW limit, 
causes that the transverse components of the closed string momentum depend 
on  $\l$ as   
 $p_i\sim \l^{-1}$. So in this limit the Wilson operator \reef{wwone} is 
independent of $\l$ (recall that $k^a=-p^a$ and $k^i=p^i$), and the 
equation \reef{tone} reduces to
\beqa
T_{\phi}(y)&=&1\,\,,\nonumber\\
T_h^{ab}(y)&=&G^{ab}
- 2G^{ca}\theta^{bd} \hF_{dc}-\theta^{ca}\theta^{bd}D_d\hP_i D_c\hP^i
+\theta^{ca}\theta^{bd}\hF_{de}G^{ef}\hF_{fc}\,\,,
\labell{ttwo}
\nonumber\\
T_h^{ai}(y)&=&\l G^{ba}D_b\hP^i-\l \theta^{ba}D_c\hP^i G^{cd}\hF_{db}+i\l
\theta^{ab}[\hP^i,\hP^j]_MD_b\hP_j\,\,,\nonumber\\
T_h^{ij}(y)&=&\l^2 G^{ab}D_a\hP^i D_b\hP^j+\l^2[\hP^i,\hP^k]_M[\hP_k,\hP^j]_M
\,\,.
\eeqa
The equation \reef{s1one} with the above tensors and the previous 
footnote,  reproduces exactly  the result of the disk amplitude of one 
closed and infinite number of open superstrings in the SW limit 
\cite{yh1,yh2}\footnote{It was argued in \cite{yh1} that  
the on-shell condition on the closed string momentum, in the SW limit, 
causes  the momentum  in the non-commutative directions  not to be arbitrary. 
Hence, one can not perform the inverse Fourier transform of \reef{s1one} in 
those directions.}.
Note that in the limit $\theta\rightarrow 0$ the result in \reef{tone} 
reduces to the energy-momentum tensor of the commutative theory.

\subsection{Non-commutative theory in  terms of commutative fields}

We have seen that, using the SW map with one open Wilson contour,  
the action \reef{biact} in the 
presence of the background B-flux can be mapped into the  action \reef{sone} 
that is
 in terms of non-commutative 
fields.  Even though the commutative action \reef{biact} 
is valid for the slowly varying open string fields, the fields in the 
non-commutative  action \reef{sone} need not to be slowly varying fields. 
Comparing this action with the string theory calculations \cite{yh1,yh2}, 
one concludes that there is no more derivative correction term to this 
action in the SW limit. On the other hand,
the commutative action \reef{biact}, can not describe 
properly the string theory in the SW limit. The derivative of $F_{ab}$ 
 and the second derivative of the scalar fields that 
are not included in it may have significant effects\footnote{Note 
that the operation in the step 4 in section 2 which reduces the 
infinite number of Wilson lines to only one line, causes that the 
commutative action \reef{biact} and the non-commutative action \reef{sone} 
not to be identical.}. However, when these derivative terms  are small 
compared to $\l$, one may use the commutative action \reef{biact}  
for describing, in the SW limit, the non-commutative theory.
When there is no closed string field, the action \reef{biact}  in the SW 
limit reduces to the following commutative U(1) gauge theory:
\beqa
\hS&=&-\frac{1}{g_{YM}^2}\int d^{p+1}x\,\sqrt{-\det G}\frac{1}
{\sqrt{\det(\cF)}}\left(-\frac{1}{4}\cF^a{}_b G^{bc}F_{cd}
\cF^d{}_e G^{ef}F_{fa}\right.\nonumber\\
&&+\frac{1}{2}\cF^a{}_b G^{bc}\prt_c\Phi^i\prt_a\Phi_i
-\frac{1}{2}\cF^a{}_b\theta^{bc}\prt_c\Phi^i\prt_d\Phi_i\cF^d{}_e 
G^{ef}F_{fa}\nonumber\\
&&\left.-\frac{1}{4}\cF^a{}_b\theta^{bc}\prt_c\Phi^i\prt_d
\Phi_i\cF^d{}_e\theta^{ef}\prt_f\Phi^j\prt_a\Phi_j+\cdots\right)\,\,,
\labell{sthree}
\eeqa
where dots represent the derivative terms, and we have defined
\beqa
\cF^a{}_b&=&(\frac{1}{1+\theta F})^a{}_b\,\,.
\nonumber
\eeqa
In expanding the square root in the action \reef{biact} around the 
background fields, one finds also a terms of order $\l^{-2}$ which 
is $(\l^2\sqrt{\det(\cF)})^{-1}$. However, by expanding this term, one 
can verify that it produces total derivative terms. For the special case 
that there is no background B-flux, $\cF^a{}_b=\delta^a{}_b$, and so the 
action \reef{sthree} reduces to the commutative YM theory. 
Expanding the action to linear order of $\theta^{ab}$, one finds 
the deformed abelian YM action found in \cite{abelian}\footnote{The 
fact that the non-commutative super YM theory can be written in terms 
of the 
commutative BI action up to the derivative correction terms, 
was used in \cite{paolo} to study some symmetry 
of the non-commutative theory.}.  

The evaluation of the linear couplings of the dilaton and the graviton 
to the non-commutative D-brane in terms of the commutative fields is also 
straightforward. The results are
\beqa
S_{\phi}&=&\frac{2\ka}{g_{YM}^2\l^2}\int\frac{ d^{10}p}{(2\pi)^{10}}\,
\sqrt{-\det G}\left(\hat{\phi}'(p)\hat{T}_{\phi}(p)\right)\,\,,\nonumber\\
S_h&=&-\frac{\ka}{g_{YM}^2\l^2}\int \frac{d^{10}p}{(2\pi)^{10}}\,
\sqrt{-\det G}
\left(\hat{h}_{\mu\nu}(p)\hat{T}_h^{\mu\nu}(p)\right)\,\,,
\nonumber
\eeqa
where
\beqa
\hat{T}_{\phi}(p)&=&\int d^{p+1}y\, T_{\phi}(y^a)
e^{-ip_ay^a+i\l p_i\Phi^i}\,\,,\nonumber\\
\hat{T}^{\mu\nu}_h(p)&=&\int d^{p+1}y \,T_h^{\mu\nu}(y^a)
e^{-ip_ay^a+i\l p_i\Phi^i}\,\,,
\nonumber
\eeqa
and
\beqa
T_{\phi}&=&\frac{1}{\sqrt{\det(\cF)}}(1+\cdots)\,\,,\nonumber\\
T_h^{ab}&=&\frac{1}{\sqrt{\det(\cF)}} 
\left(\cF^a{}_c G^{cb}-\cF^a{}_c(\theta^{cd}\prt_d
\Phi^i\prt_e\Phi_i+G^{cd}F_{de})\cF^e{}_f\theta^{fb}+\cdots\right)\,\,,
\nonumber\\
T_h^{ai}&=&\frac{\l}{\sqrt{\det(\cF)}}
\left(\cF^a{}_bG^{bc}\prt_c\Phi^i-
\cF^{(a}{}_c(\theta^{cd}\prt_d\Phi^j\prt_e\Phi_j+
G^{cd}F_{de})\cF^e{}_f\theta^{fg)}\prt_g\Phi^i+\cdots
\right)\,\,,\nonumber\\
T_h^{ij}&=&\frac{\l^2}{\sqrt{\det(\cF)}}\left(\cF^a{}_b G^{bc}
\prt_c\Phi^i\prt_a\Phi^j-\cF^a{}_b(\theta^{bc}
\prt_c\Phi^k\prt_d\Phi_k+G^{bc}F_{cd})\cF^d{}_e\theta^{ef}
\prt_f\Phi^i\prt_a\Phi^j+\cdots\right)\,\,.
\nonumber
\eeqa
Here again the dots represents the derivative terms.
Since the transverse components of the  closed string fields depend 
on $\l$ as   $p_i\sim\l^{-1}$, the exponential factors above are independent 
of $\l$.
 One expects that, when the B-flux is zero, the above results and the result 
in \reef{sthree} reduce to the 
results in \reef{ttwo} and in \reef{stwo} for  B=0.  So we recover the known 
fact that the  derivative correction 
terms to the ordinary DBI action
are such that, in the SW limit and for  B=0, they all vanish.

In writing the abelian action \reef{sthree} and the energy-momentum 
tensors above, we have taken the SW limit of  the abelian DBI action 
\reef{biact}. Extension of these results to the non-abelian case is 
straightforward. One should take the SW limit of the non-abelian extension 
of the DBI action\cite{rm1}.
 
\section{Discussion}

In this paper we have found an action for transformation of the ordinary 
DBI action, which includes  the non-constant massless closed string  
fields,  
under the SW map with one open Wilson contour. 
This action contains  derivatives of the 
non-commutative open string  fields through the $*_N$-product,
 and  one  Wilson operator. 
The world-volume indices of the open string fields are raised by the open 
string metric $G^{ab}$. They all stems from the SW map. The action also 
inherits the non-constant closed string fields from the commutative action.  
We have shown that the action is invariant under the non-commutative 
gauge transformation, and it is consistent with the T-duality rules. 
>From the non-constant closed string fields, we have found the linear 
couplings of the dilaton and the graviton to the non-commutative gauge 
fields for arbitrary $\l$ and background fields. In the SW limit, we 
have shown that these couplings reduce to the known results of the 
energy-momentum tensor of the non-commutative super YM theory. Hence, we 
reached to the conclusion that the non-commutative  action \reef{sone}
in the SW limit has no further derivative correction terms.

In sect. 5,  we compared the proposed action \reef{sone} with the 
disk S-matrix elements of one closed string and infinite number of the open 
string states in the SW limit,  and found  exact agreement 
between them. In this limit,  the background fields are restricted to 
$g_{ai}=B_{ai}=B_{ij}=0$. Using  the action \reef{sone} it is easy to extend 
the results in section 5 to include these background fields as well. It would 
be interesting then to release that restriction in the disk amplitude 
calculation and check  if still one finds agreement between the string 
theory calculation and the action \reef{sone}. 

The  recipe  for transforming ordinary $D_9$-brane action under SW map 
with one Wilson line in sect. 2 includes the transformations like 
$\left(f_1(f_2f_3)*_2(f_4f_5f_6)*_3\cdots f_N\right)
\rightarrow (f_1f_2\cdots f_N)*_N$, where 
$f_1,\cdots f_N$ are  any of $\hA^a$ or $\hF_{ab}$. Using the results in 
sect. 3, this  means that we impose an extra operation that reduces the 
number of the open Wilson contours, $n$, to one contour, 
\ie $L_*^n\rightarrow L_*$. This stems from the fact that the world-sheet 
corresponding to the tree level effective action (disk) has one boundary. 
For one loop level effective action, the world-sheet (cylinder) has 
two boundaries. So in that case, one should impose an operation that 
reduces the number of the Wilson lines  to two. It would be interesting, 
then, to impose this construction to the one loop effective action of 
the commutative theory to find the one loop effective action of the 
non-commutative theory\cite{liu}.  

In this paper, we have found the non-commutative action for $D_p$-brane by 
applying the familiar rules of T-duality on the $D_9$-brane. One may wish 
to find this action directly from the commutative $D_p$-brane action and 
the prescriptions in sect.2. In this way the prescription is as follows:
\\
1.  Start with commutative $D_p$-brane.
\\
2. Expand the square root for quantum fluctuations 
around the background $g_{ab}+\l B_{ab}$ field. 
\\
3. Transform each commutative $F_{ab}$, $\prt_a\Phi^i$ and $\Phi^i$  to 
their non-commutative counterparts  according to the SW map.  
 The SW map for $\prt_a\Phi^i$ and $\Phi^i$ can 
be read from the corresponding map for $F_{ab}$ and $A_a$ by dimensional 
reduction \cite{mg2,mg1,sm1}. 
\\
4. Reduce the number of Wilson lines to only one line, \ie 
 replace the $*_N$ between $\hF_{ab}$, $D_a\hP^i$, 
and  any other field, and then simplify the result using the identity 
\reef{idenone}. 

Let us work the above prescription for one simple example. Consider the 
coupling of one Kalb-Ramond closed string field with two transverse scalar 
fields, that is
\[
V^{ab}b_{ij}\prt_a\Phi^i\prt_b\Phi^j=\frac{1}{\l}b_{ij}\theta^{ab}
\prt_a\Phi^i\prt_b\Phi^j\,\,.
\]
According to the above prescription it transforms to 
\beqa
\frac{1}{\l}b_{ij}\theta^{ab}\left(D_a\hP^i D_b\hP^j\right)*_2+
\cdots&=&\frac{i}{\l}b_{ij}[\hP^i,\hP^j]_M+\cdots\,\,,
\nonumber
\eeqa
where we have used the identity $\theta^{ab}(\prt_a f\prt_b g)*_2=i[f,g]_M$. 
The above  term is exactly reproduced by the second square root in 
\reef{lllone}. It is not reproduced by the first square root because of the 
rule in  \reef{vtog}. This term is also reproduced by the disk S-matrix 
elements of one closed and two open string states\cite{mine}. One needs  
more complicated identities to produce the terms in the action \reef{lllone} 
which have more than one commutator.
Alternatively, one may compare the terms coming from the above prescription 
and the non-commutative action \reef{lllone} to find the more complicated 
identities between the $*_N$.
As an  example consider the following term which stems from the expansion 
of ordinary DBI action: 
\beqa
V^{ab}\prt_b\Phi^i\prt_c\Phi_iV^{cd}\prt_d\Phi^j\prt_a\Phi_j&=&
G^{ab}\prt_b\Phi^i\prt_c\Phi_i G^{cd}\prt_d\Phi^j\prt_a\Phi_j\nonumber\\
&&+\frac{1}{\l^2}\theta^{ab}\prt_b\Phi^i\prt_c\Phi_i\theta^{cd}
\prt_d\Phi^j\prt_a\Phi_j\,\,.\nonumber
\eeqa
According to the above prescription they transform to
\beqa
\left(G^{ab}D_b\hP^iD_c\hP_iG^{cd}D_d\hP^j D_a\hP_j\right)*_4
+\frac{1}{\l^2}(\theta^{ab}D_b\hP^iD_c\hP_i\theta^{cd}D_d\hP^jD_a\hP_j)*_4+
\cdots\,\,.
\labell{iden}
\eeqa
However the corresponding terms from the action \reef{lllone} is
\beqa
\left(G^{ab}D_b\hP^iD_c\hP_iG^{cd}D_d\hP^j D_a\hP_j\right)*_4
-\frac{1}{\l^2}[\hP_j,\hP_i]_M*_2[\hP^j,\hP^i]_M+\cdots\,\,.
\nonumber
\eeqa
The first and second terms above are reproduced by the expansion of the 
first and the second determinant in the square  root in \reef{lllone}, 
respectively.  
Comparing them  with \reef{iden}, one would expect the following 
identity
\beqa
(\theta^{ab}\prt_af_1\prt_bg_1\theta^{cd}\prt_cf_2\prt_d g_2)*_4&=&-
[f_1,g_1]_M*_2[f_2,g_2]_M\,\,.
\nonumber
\eeqa
Examining the higher order terms, one would expect the following 
general identity
\beqa
(\theta^{a_1b_1}\prt_{a_1}f_1\prt_{b_1}g_1\cdots \theta^{a_nb_n}
\prt_{a_n}f_n\prt_{b_n}g_n)*_{2n}&=&(i)^n([f_1,g_1]_M
\cdots [f_n,g_n]_M)*_n\,\,.
\nonumber
\eeqa
One may try to prove this recursion formula between $*_{2n}$ and $*_{n}$ 
using the definition of $*_N$-product\cite{liu,mg1,yk}.

The commutative action \reef{biact} includes, among other things, 
the couplings of the closed string states to commutative gauge fields 
when derivative of $F_{ab}$ is small compare to $\l$. Whereas, 
the action \reef{lllone} or 
\reef{sone} includes the similar couplings to non-commutative gauge 
fields with no restriction on $\hF_{ab}$ in the SW limit. When the 
derivative of $\hF_{ab}$ is small, one can replace the $*_N$-product 
with the ordinary product. In that case, the operation in the step 4 
in the section 2 is identity operator. So one expects that the commutative 
and non-commutative actions to be identical. Therefore, one may compare 
the closed string couplings in these actions to extract some 
transformation between the 
commutative and the  non-commutative fields under the SW map. 
For example, consider  linear 
coupling of the dilaton  to commutative gauge field on the world-volume of 
$D_9$-brane \reef{biact}, that is
\beqa
S&=&T_9\sqrt{-\det(g_{\mu\nu}+\l B_{\mu\nu})}\int d^{10}x\,\phi
\sqrt{\det(1+\l G F+\theta F)}\,\,,
\nonumber
\eeqa
where the $g_{\mu\nu}+\l B_{\mu\nu}$ is the constant background field. 
Now the same thing in terms of non-commutative fields is
\beqa
\hS&=&T_9\sqrt{-\det(g_{\mu\nu}+\l B_{\mu\nu})}\int d^{10}x\,
\left(\phi(\hA)\sqrt{\det(1+\l G \hF)}\right)\,\,,
\nonumber
\eeqa
where we have used the prescription \reef{vtog}. Using the manipulation in 
sect. 3, one will find the following transformation
\beqa
\sqrt{\det(1+\l G F+\theta F)}(k)&=&\int d^{10}y\,
\left[W(k^{\mu},y^{\mu},C)\sqrt{\det(1+\l G\hF)}\right]e^{ik\cdot y}\,\,
\nonumber\\
&=&\int d^{10}y\,
\sqrt{\det(1+\l G\hF)}e^{ik\cdot (y+\theta \hA)}\,\,.
\nonumber
\eeqa
Now we change to the new coordinate\cite{liu}
\beqa
X(y)&=&y+\theta\hA(y)\,\,\,\,\,;\,\,\,\,\,\int 
d^{10}y\longrightarrow\int \frac{d^{10}X}{\sqrt{\det(1-\theta\hF)}}\,\,.
\nonumber
\eeqa
>From this we find that
\beqa
\sqrt{\det(1+\l G F+\theta F)}(X)=\frac{\sqrt{
\det(1+\l G\hF)}}{\sqrt{\det(1-\theta \hF)}}\,\,.
\nonumber
\eeqa
Doing some simple algebra, one finds
\beqa
F_{\mu\nu}(X(y))&=&\left(\hF(y)\frac{1}{1-\theta\hF(y)}\right)_{\mu\nu}(y)\,\,.
\nonumber
\eeqa
This is exactly the identity found in \cite{liu}. In \cite{liu} above 
identity was derived from the conjectured exact solution  to the SW 
differential equation. That conjecture was then confirmed in \cite{yh3,sm1}. 
The solution to the SW 
differential equation found in \cite{liu,yh3,sm1} is valid only for $U(1)$ 
gauge theory. The reason is that  if one directly integrates the SW 
differential equation for the case of non-abelian gauge fields, one will 
find that a new type of multiplication, the one which  is not the 
$*_N$-product, appears in the solution\cite{mine}. So it would be 
interesting, using \cite{liu},  to extend the perturbative solution 
found in \cite{mine} to an 
exact solution for the non-abelian cases. Other progress in this 
direction has been made in \cite{jurco,brace}.

We have seen that starting from the ordinary DBI action and mapping  
it to the non-commutative variables smeared along one Wilson line 
and then taking the SW limit,  one 
finds exactly the known result in the  non-commutative super YM theory, 
\eg the energy-momentum tensor. However, the energy-momentum tensor for 
the D-branes of the bosonic theory  is not the same as for the D-branes of 
the superstring theory\cite{yh1,yh2}. It would be interesting then to find 
a commutative action  that reproduces the result for the bosonic case 
using the prescription given in this paper.
  
The Matrix theory is a special limit of the non-abelian non-commutative 
super YM theory. The dimensional reduction of the YM theory along the 
world-volume space, reduces it to the Matrix theory. In this case, the SW 
limit reduces to the DKPS limit\cite{mrd1}. As we discussed in the text, 
to  extend  the non-commutative action \reef{sone} to the non abelian case 
one should change the fields to the matrix valued fields, and at the same 
time should use the symmetrized trace prescription. Hence, the known form of 
the Matrix theory action is the dimensional reduction of the non-abelian form 
of the action \reef{stwo}, that is,
\beqa
S&=&-\frac{1}{g_{YM}^2}\int dt \,\sqrt{-\det G}\,\,{\rm Tr}\left( 
\frac{1}{2}G^{00}D_0\Phi_i D_0\Phi^i
+\frac{1}{4}[\Phi^I,\Phi^J][\Phi_J,\Phi_I]\right)\,\,,
\nonumber
\eeqa
where now the indices $I,J=1,2,\cdots,9$, and there is no non-commutative 
parameter. The energy-momentum tensor of the theory can be read from the 
results in \reef{s1one}--\reef{tone}, that is,
\beqa
S_{\phi}&=&\frac{2\ka}{g_{YM}^2\l^2}\int\frac{d^{10}p}{(2\pi)^{10}}
\sqrt{-\det G}\left(\hat{\phi}'(p)\hat{T}_{\phi}(p)\right)\,\,,\nonumber\\
S_h&=&-\frac{\ka}{g_{YM}^2\l^2}\int\frac{d^{10}p}{(2\pi)^{10}}\sqrt{-\det G}
\left(\hat{h}_{\mu\nu}(p)\hat{T}_{h}^{\mu\nu}(p)\right)\,\,,
\nonumber
\eeqa
where
\beqa
\hat{T}_{\phi}(p)&=&\int dt \,\,\Tr\left(T_{\phi}(t)e^{-ip_0t+i\l 
p_I\Phi^I}\right)\,\,,\nonumber\\
\hat{T}^{\mu\nu}_h(p)&=&\int dt\,\, {\rm STr}\left(T_h^{\mu\nu}(t)
e^{-ip_0t+i\l p_I\Phi^I}\right)\,\,,
\nonumber
\eeqa
and 
\beqa
T_{\Phi}(t)&=&1\,\,,\nonumber\\
T^{00}_h(t)&=&G^{00}\,\,,\nonumber\\
T^{0I}_h(t)&=&\l G^{00}D_0\Phi^I\,\,,\nonumber\\
T^{IJ}_h(t)&=&\l^2 G^{00}D_0\Phi^I D_0\Phi^J+
\l^2[\Phi^I,\Phi^K][\Phi_K,\Phi^J]\,\,.\nonumber
\eeqa
These expressions are exactly the known  results for the energy-momentum 
tensor of the Matrix theory\cite{dk,wi,yh2}. Note that in this case there
is no non-commutative parameter, no Wilson line and no path ordering 
prescription. Therefore, we have used the symmetrized trace 
prescription instead of taking the trace of the path ordered terms.

{\bf Acknowledgments}

I would like to acknowledge useful conversation with F. Ardalan. 
This work was supported by University of Birjand and IPM.

\end{document}